\documentclass{patmorin}

\usepackage{amsmath,amsfonts,amsthm}

\newcommand{\eg}{\textsl{e.g.}}
\newcommand{\ie}{\textsl{i.e.}}
\newcommand{\BigOh}[1]{O\!\left(#1\right)}
\newcommand{\LittleOh}[1]{o\!\left(#1\right)}

\newcommand{\BigTheta}[1]{\Theta\!\left(#1\right)}
\newcommand{\U}{\mathcal{U}}

\newtheorem{theorem}{Theorem}{\bfseries}{\itshape}

\newcommand{\keywords}[1]{\noindent\textbf{Keywords:} #1}

\begin{document}

\title{\MakeUppercase{Biased Predecessor Search}\thanks{Partially
supported by the Danish Council for Independent Research,
Natural Sciences, grant DFF-1323-00247.}}
\author{%
Prosenjit~Bose,\thanks{\affil{Carleton University},
\email{\{jit,jhowat,morin\}@scs.carleton.ca}}\:\:
Rolf~Fagerberg,\thanks{\affil{University of Southern Denmark},
  \email{rolf@imada.sdu.dk}}\:\:
John~Howat,\footnotemark[2]\:\:
and Pat~Morin\footnotemark[2]
}
\date{}

\maketitle

\begin{abstract}
We consider the problem of performing predecessor searches in a bounded universe while achieving query times that depend on the distribution of queries. We obtain several data structures with various properties: in particular, we give data structures that achieve expected query times logarithmic in the entropy of the distribution of queries but with space bounded in terms of universe size, as well as data structures that use only linear space but with query times that are higher (but still sublinear) functions of the entropy. For these structures, the distribution is assumed to be known. We also consider individual query times on universe elements with general weights, as well as the case when the distribution is not known in advance.

\end{abstract}

\section{Introduction}
\label{:introduction}

The notion of \emph{biased searching} has received significant attention
in the literature on ordered dictionaries. In ordered dictionaries, the
central operation is predecessor queries---that is, queries for the
largest element stored in the data structure that is smaller than a
given query value. The setting is biased when each element $i$ of the
data structure has some probability $p_i$ of being queried, and we wish
for queries to take a time related to the inverse of the probability of
that query. For example, a \emph{biased search tree} \cite{biasedsearch}
can answer a query for item $i$ in time $\BigOh{\log
1/p_i}$.\footnote{In this paper, we define $\log x = \log_2 (x + 2)$.}
For biased predecessor queries, also the gaps between consecutive
elements of the data structure are assigned probabilities of being
searched for \cite[p.~564]{optimumbst,biasedsearch}. Recall that $\sum_i
p_i \log (1/p_i)$ is the \emph{entropy} of the distribution of
queries. In terms of this quantity, we note that the \emph{expected}
query time in a biased search tree is linear in the entropy of the query
distribution, and that this is optimal for binary search
trees~\cite[Thm.~A]{biasedsearch}.\footnote{As will be apparent from our
results, in bounded universes this lower bound does not hold, and one can achieve query times below it.}

Binary search trees work in the \emph{comparison-based} setting where keys only can be accessed by comparisons. Predecessor searches have also been researched extensively in the context of \emph{bounded universes} where keys are integers of bounded range whose bits may be accessed individually. More precisely, let $\U = \{ 0,1,\ldots,U-1 \}$ be the universe of possible keys, and consider a static subset $S = \{ s_1, s_2, \ldots, s_n \} \subseteq \U$, where $s_1 < s_2 < \cdots < s_n$. Predecessor searches in this context admit data structures with query times that are not only a function of $n$, but also of $U$. For example, van Emde Boas trees \cite{veb} can answer predecessor queries in time $\BigOh{\log \log U}$.

A natural question---but one which has been basically unexplored---is how to combine these two areas of study to consider biased searches in bounded universes. In this setting, we have a probability distribution $D = \{ p_0, p_1, \ldots, p_{U-1} \}$ over the universe $\U$ such that the probability of receiving $i \in \U$ as a query is $p_i$ and $\sum_{i=0}^{U-1} p_i = 1$. We wish to preprocess $\U$ and $S$, given $D$, such that the time for a query is related to $D$.

The motivation for such a goal is the following. Let $H = \sum_{i=0}^{U-1} p_i \log (1/p_i)$ be the entropy of the distribution $D$. Recall that the entropy of a $U$-element distribution is between $0$ and $\log U$. Therefore, if an expected query time of $\BigOh{\log H}$ can be achieved, this for any distribution will be at most $\BigOh{\log \log U}$, which matches the performance of van Emde Boas trees \cite{veb}. However, for lower-entropy distributions, this will be faster---as a concrete example, an exponential distribution (say, $p_i = \Theta(1/2^i)$) has $H = O(1)$ and will yield support of queries in expected constant time. In other words, such a structure will allow bias in the query sequence to be exploited for ordered dictionaries over bounded universes. Hence, perhaps the most natural way to frame the line of research in this paper is by analogy: the results here are to biased search trees as van Emde Boas trees (and similar structures) are to binary search trees.

\paragraph{Our results.} The results presented here can be divided into four categories. In the first we give two variants of a data structure that obtains $\BigOh{\log H}$ query time but space that is bounded in terms of $U$. In the second we give a solution that obtains space that is linear in $n$ but has query time $\BigOh{\sqrt{H}}$. In bounded universe problems, $n$ is always smaller than $U$ (often substantially so), so these two categories can be seen as representing a time-space trade-off. In the third we consider individual query times on universe elements with general weights. In the fourth we consider query times related to the working-set number (which is defined as the number of distinct predecessors reported since the last time a particular predecessor was reported), so that the query distribution need not be known in advance. Our methods use hashing and existing (unbiased) predecessor structures for bounded universes~\cite{bf,yfast} as building blocks.

\paragraph{Organization.} The rest of the paper is organized in the following way. We first complete the current section by reviewing related work. In Section~\ref{:goodquery} we show how to obtain good query times at the expense of large space. In Section~\ref{:goodspace} we show how to obtain good space at the expense of larger query times. We conclude in Section~\ref{:conclusion} with a summary of the results obtained and possible directions for future research.

\subsection{Related Work}
\label{:introduction:related}

It is a classical result that predecessor searches in bounded universes can be performed in time $\BigOh{\log \log U}$. This was first achieved by van Emde Boas trees \cite{veb}, and later by $y$-fast tries \cite{yfast}, and Mehlhorn and N{\"a}her \cite{bounded}. Of these, van Emde Boas trees use $\BigOh{U}$ space, while the other two structures use $\BigOh{n}$ space.

These bounds can be improved to

\begin{displaymath}
\BigOh{ \min\left\{ \frac{\log \log U}{\log \log \log U}, \sqrt{\frac{\log n}{\log \log n}} \right\} }
\end{displaymath}
using $n^{\BigOh{1}}$ space \cite{bf}. By paying an additional $\BigOh{\log \log n}$ factor in the first half of this bound, the space can be improved to $\BigOh{n}$ \cite{bf}. P\u{a}tra\c{s}cu and Thorup later effectively settled this line of research with a set of time-space trade-offs \cite{pt}.

Departing the bounded universe model for a moment and considering only biased search, perhaps the earliest such data structure is the optimum binary search tree \cite{optimumbst}, which is constructed to be the best possible static binary search tree for a given distribution. Optimum binary search trees take a large amount of time to construct; in linear time, however, it is possible to construct a binary search tree that answers queries in time that is within a constant factor of optimal \cite{nearlyoptimalbst}. Even if the distribution is not known in advance, it is still possible to achieve the latter result (\eg, \cite{unified,splay}).

Performing biased searches in a bounded universe is essentially unexplored, except for the case where the elements of $S$ are drawn from $D$ rather than the queries \cite{randominput}. In that result, $D$ need not be known, but must satisfy certain smoothness constraints, and a data structure is given that supports $\BigOh{1}$ query time with high probability and $\BigOh{\sqrt{\log n / \log \log n}}$ worst-case query time, using $\BigOh{n^{1+\epsilon}}$ bits of space, which can be reduced to $\BigOh{n}$ space at the cost of a $\BigOh{\log \log n}$ query time (with high probability). It is worth noting that this data structure is also dynamic.

A related notion is to try to support query times related to the distribution in a less direct way. For example, \emph{finger searching} can be supported in time $\BigOh{\sqrt{\log d / \log \log d}}$ where $d$ is the number of keys stored between a \emph{finger} pointing at a stored key and the query key \cite{fingerram}. There is also a data structure that supports such searches in expected time $\BigOh{\log \log d}$ for a wide class of input distributions \cite{randominputfinger}. Finally, a query time of $\BigOh{\log \log \Delta}$, where $\Delta$ is the difference between the element queried and the element returned, can also be obtained \cite{local}.

Other problems in bounded universes can also be solved in similar ways. A \emph{priority queue} that supports insertion and deletion in time $\BigOh{\log \log d'}$, where $d'$ is the difference between the successor and predecessor (in terms of priority) of the query, is known \cite{priorityqueue}, as well as a data structure for the \emph{temporal precedence problem}, wherein the older of two query elements must be determined, that supports query time $\BigOh{\log \log \delta}$, where $\delta$ is the temporal distance between the given elements \cite{temporal}.

\section{Supporting $\BigOh{\log H}$ Query Time}
\label{:goodquery}

In this section, we describe how to achieve query time $\BigOh{\log H}$
using space that is bounded in terms of $U$.

\subsection{Using $\BigOh{n + U^\epsilon}$ Space}
\label{:goodquery:bigspace}

Let $\epsilon > 0$. We place all elements $i \in \U$ with probability $p_i \ge (1/U)^\epsilon$ into a hash table $T$, and with each element we store a pointer to its predecessor in $S$ (which never changes since $S$ is static). All elements of $S$ are also placed into a $y$-fast trie over the universe $\U$. Since there are at most $U^\epsilon$ elements with probability greater than $(1/U)^\epsilon$, it is clear that the hash table requires $\BigOh{U^\epsilon}$ space. Since the $y$-fast trie requires $\BigOh{n}$ space, we have that the total space used by this structure is $\BigOh{n + U^\epsilon}$. To execute a search, we check the hash table first. If the query (and thus the answer) is not stored there, then a search is performed in the $y$-fast trie to answer the query.

The expected query time is thus
\begin{eqnarray*}
&& \sum_{i \in T} p_i \BigOh{1} + \sum_{i \in \U \setminus T} p_i \BigOh{\log \log U}\\
&=& \BigOh{1} + \sum_{i \in \U \setminus T} p_i \BigOh{\log \log U} \\
&=& \BigOh{1} + \sum_{i \in \U \setminus T} p_i \BigOh{\log \log \left((U^\epsilon)^{1/\epsilon}\right)} \\
&=& \BigOh{1} + \sum_{i \in \U \setminus T} p_i \BigOh{\log ((1/\epsilon) \log U^\epsilon)} \\
&=& \BigOh{1} + \sum_{i \in \U \setminus T} p_i \BigOh{\log (1/\epsilon)} + \sum_{i \in \U \setminus T} p_i \BigOh{\log \log U^\epsilon} \\
&=& \BigOh{1} + \BigOh{\log(1/\epsilon)} + \sum_{i \in \U \setminus T} p_i \BigOh{\log \log \frac{1}{1/U^\epsilon}} \\
&\le& \BigOh{1} + \BigOh{\log(1/\epsilon)} + \sum_{i \in \U \setminus T} p_i \BigOh{\log \log (1/p_i)}
\end{eqnarray*}
The last step here follows from the fact that, if $i \in \U \setminus T$, then $p_i \le (1/U)^\epsilon$, and so $1/(1/U)^\epsilon \le 1/p_i$. Recall Jensen's inequality, which states that for concave functions $f$, $E[f(X)] \le f(E[X])$. Since the logarithm is a concave function, we therefore have
\begin{displaymath}
\sum_{i \in \U \setminus T} p_i \BigOh{\log \log (1/p_i)} \le \log \sum_{i \in \U \setminus T} p_i \BigOh{\log (1/p_i)} \le \BigOh{\log H}
\end{displaymath}
therefore, the expected query time is $\BigOh{\log (1/\epsilon)} + \BigOh{\log H} = \BigOh{\log (H/\epsilon)}$.

\begin{theorem}
\label{:theorem:log-h-over-epsilon}
Given a probability distribution with entropy $H$ over the possible queries in a universe of size $U$, it is possible to construct a data structure that performs predecessor searches in expected time $\BigOh{\log (H/\epsilon)}$ using $\BigOh{n + U^\epsilon}$ space for any $\epsilon > 0$.
\end{theorem}

Theorem~\ref{:theorem:log-h-over-epsilon} is a first step towards our goal. For $\epsilon = 1/2$, for example, we achieve $\BigOh{\log H}$ query time, as desired, and our space usage is $\BigOh{n} + \LittleOh{U}$. This dependency on $U$, while sublinear, is still undesirable. In the next section, we will see how to reduce this further.

\subsection{Using $\BigOh{n + 2^{\log^{\epsilon} U}}$ Space}
\label{:goodquery:lessbigspace}

To improve the space used by the data structure described in Theorem~\ref{:theorem:log-h-over-epsilon}, one observation is that we can more carefully select the threshold for ``large probabilities'' that we place in the hash table. Instead of $(1/U)^\epsilon$, we can use $(1/2)^{\log^{\epsilon} U}$ for some $0 < \epsilon < 1$. The space used by the hash table is thus $\BigOh{2^{\log^{\epsilon} U}}$, which is $\LittleOh{U^\epsilon}$ for any $\epsilon > 0$. The analysis of the expected query times carries through as follows
\begin{eqnarray*}
\sum_{i \in T} p_i \, \BigOh{1} + \sum_{i \in \U \setminus T} p_i \BigOh{\log \log U}
&=& \BigOh{1} + \sum_{i \in \U \setminus T} p_i \BigOh{\log \log U} \\
&=& \BigOh{1} + \sum_{i \in \U \setminus T} p_i \epsilon(1/\epsilon) \BigOh{\log \log U} \\
&=& \BigOh{1} + \sum_{i \in \U \setminus T} p_i (1/\epsilon) \BigOh{\log \left(\log^{\epsilon} U\right)} \\
&=& \BigOh{1} + \sum_{i \in \U \setminus T} p_i (1/\epsilon) \BigOh{\log \log \left( 2^{\log^{\epsilon} U} \right)} \\
&\le& \BigOh{1} + \sum_{i \in \U \setminus T} p_i (1/\epsilon) \BigOh{\log \log (1/p_i)} \\
&\le& \BigOh{1} + (1/\epsilon) \sum_{i \in \U \setminus T} p_i \BigOh{\log \log (1/p_i)} \\
&\le& \BigOh{(1/\epsilon) \log H}
\end{eqnarray*}

\begin{theorem}
\label{:theorem:1/e-log-h}
Given a probability distribution with entropy $H$ over the possible queries in a universe of size $U$, it is possible to construct a data structure that performs predecessor searches in expected time $\BigOh{(1/\epsilon) \log H}$ using $\BigOh{n + 2^{\log^{\epsilon} U}}$ space for any $0 < \epsilon < 1$.
\end{theorem}

\subsection{Individual Query Times for Elements}
\label{:goodquery:individual}

Observe that part of the proof of Theorem~\ref{:theorem:1/e-log-h} is to show that an individual query for an element $i \in \U \setminus T$ can be executed in time $\BigOh{(1/\epsilon) \log \log 1/p_i}$ time.  Since the query time of elements in $T$ is $\BigOh{1}$, the same holds for these. More generally, the structure can support arbitrarily weighted elements in~$\U$. Suppose each element $i \in \U$ has a real-valued weight $w_i > 0$ and let $W = \sum_{i=0}^{U-1} w_i$. By assigning each element probability $p_i = w_i/W$, we achieve an individual query time of $\BigOh{(1/\epsilon) \log \log (W/w_i)}$, which is analogous to the $\BigOh{\log W/w_i}$ query time of biased search trees~\cite{biasedsearch}. Since the structure is static, we can use perfect hashing for the hash tables involved ($T$ as well as those in the $y$-fast trie), hence the search time is worst-case.

\begin{theorem}
\label{:theorem:weighted}
Given a positive real weight $w_i$ for each element $i$ in a universe of size $U$, such that the sum of all weights is $W$, it is possible to construct a data structure that performs a predecessor search for item $i$ in worst-case time $\BigOh{(1/\epsilon) \log \log (W/w_i)}$ using $\BigOh{n + 2^{\log^{\epsilon} U}}$ space for any $0 < \epsilon < 1$.
\end{theorem}

\section{Supporting $\BigOh{n}$ Space}
\label{:goodspace}

In this section, we describe how to achieve space $\BigOh{n}$ by accepting a larger query time $\BigOh{\sqrt{H}}$. We begin with a brief note concerning input entropy vs. output entropy.

\paragraph{Input vs.\ Output Distribution.} Until now, we have discussed the \emph{input} distribution, \ie, the probability $p_i$ that $i \in \U$ is the \emph{query}. We could also discuss the \emph{output} distribution, \ie, the probability $p^*_i$ that $i \in \U$ is the \emph{answer} to the query. This distribution is defined by $p^*_i = 0$ if $i \notin S$ and $p^*_i = \sum_{j=s_k}^{s_{k+1}-1} p_j$ if $i \in S = \{ s_1, s_2, \ldots, s_n \}$ with $i = s_k$.

Suppose we can answer a predecessor query for $i$ in time $\BigOh{\log \log 1/p^*_{pred(i)}}$ where $pred(i)$ is the predecessor of $i$. Then the expected query time is
\begin{displaymath}
\sum_{i \in \U} p_i \BigOh{\log \log 1/p^*_{pred(i)}}
\end{displaymath}
Since $p_i \le p^*_{pred(i)}$ for all $i$, this is at most $\sum_{i \in \U} p_i \log \log 1/p_i$, \ie, the entropy of the input distribution. It therefore suffices to consider the output distribution.

Our data structure will use a series of data structures for predecessor search~\cite{bf} that increase doubly-exponentially in size in much the same way as the working-set structure \cite{unified}. Recall from Section~\ref{:introduction:related} that there exists a linear space data structure that is able to execute predecessor search queries in time $\BigOh{\min\left\{ \frac{\log \log n \cdot \log \log U}{\log \log \log U}, \sqrt{\frac{\log n}{\log \log n}} \right\}}$ \cite{bf}. We will maintain several such structures $D_1,D_2,\ldots$, where each $D_j$ is over the universe $\U$ and stores $2^{2^j}$ elements of $S$. In more detail, sorting the elements of $S$ by probability into decreasing order, we store the first $2^{2^1}$ elements in $D_1$, the next $2^{2^2}$ elements in $D_2$, etc. In general, $D_j$ contains the $2^{2^j}$ elements of highest probability that are not contained in any $D_k$ for $k < j$. Note that here, ``probability'' refers to the \emph{output} probability.

Searches are performed by doing a predecessor search in each of $D_1,D_2,\ldots$ until the answer is found. Along with each element we store a pointer to its successor in~$S$. When we receive the predecessor of the query in $D_j$, we check its successor to see if that successor is larger than the query. If so, the predecessor in $D_j$ is also the real predecessor in~$S$ (\ie, the answer to the query), and we stop the process. Otherwise, the real predecessor in~$S$ is somewhere between the predecessor in $D_j$ and the query, and can be found by continuing to $D_{j+1},D_{j+2},\ldots$. This technique is known from~\cite{ws-implicit}.

We now consider the search time in this data structure. Suppose the
process stops by finding the correct predecessor of the query $i$ in
$D_j$ where $j > 1$ (otherwise, the predecessor was found in $D_1$ in
$\BigOh{1}$ time). $D_{j-1}$ contains $2^{2^{j-1}}$ elements all of
which have (output) probability at least $p^*_{pred(i)}$. Since the sum
of the probabilities of these elements is at most one, it follows that
$p^*_{pred(i)} \le 1/2^{2^{j-1}}$. Equivalently, $j$ is $\BigOh{\log
\log 1/p^*_{pred(i)}}$. The total time spent searching is bounded by
\begin{equation}\label{:equation:bfeq}
\sum_{k=1}^j \sqrt{\frac{\log 2^{2^k}}{\log \log 2^{2^k}}} = \sum_{k=1}^j \sqrt{\frac{2^k}{k}} = \BigOh{\sqrt{\frac{2^j}{j}}} = \BigOh{\sqrt{\log 1/p^*_{pred(i)}}}
\end{equation}
The second equality above follows because the terms of the summation
are exponentially increasing and hence the last term dominates the
entire sum. Therefore, since $p_i \le p^*_{pred(i)}$ for all $i$, the
expected query time is
\begin{displaymath}
\sum_{i \in \U} p_i \sqrt{\log 1/p^*_{pred(i)}} \le \sum_{i \in \U} p_i \sqrt{\log 1/p_i} \le \sqrt{H}
\end{displaymath}
The final step above follows from Jensen's inequality. To determine the space used by this data structure, observe that every element stored in $S$ is stored in exactly one $D_j$. Since each $D_j$ uses space linear in the number of elements stored in it, the total space usage is $\BigOh{n}$.

\begin{theorem}
\label{:theorem:linear-space}
Given a probability distribution with entropy $H$ over the possible queries in a universe of size $U$, it is possible to construct a data structure that performs predecessor searches in expected time $\BigOh{\sqrt{H}}$ using $\BigOh{n}$ space.
\end{theorem}

Observe that we need not know the exact distribution $D$ to achieve the result of Theorem~\ref{:theorem:linear-space}; it suffices to know the sorted order of the keys in terms of non-increasing probabilities.

Also observe that like in Section~\ref{:goodquery:individual}, the structure here can support arbitrarily weighted elements. Suppose each element $i \in \U$ has a real-valued weight $w_i > 0$ and let $W = \sum_{i=0}^{U-1} w_i$. By assigning each element probability $p_i = w_i/W$, we see that \eqref{:equation:bfeq} and the fact that $p_i \le p^*_{pred(i)}$ for all $i$ give the following.

\begin{theorem}
\label{:theorem:weightedbf}
Given a positive real weight $w_i$ for each element $i$ in a bounded universe, such that the sum of all weights is $W$, it is possible to construct a data structure that performs a predecessor search for item $i$ in worst-case time $\BigOh{\sqrt{\log (W/w_{i})}}$ using $\BigOh{n}$ space.
\end{theorem}

Furthermore, since the predecessor search structure used for the $D_j$'s above is in fact dynamic \cite{bf}, we can even obtain a bound similar to the working-set property: a predecessor search for item $i$ can be answered in time $\BigOh{\sqrt{\log w(i)}}$ where $w(i)$ is the number of distinct predecessors reported since the last time the predecessor of $i$ was reported. This can be accomplished using known techniques \cite{unified}, similar to the data structure of Theorem~\ref{:theorem:linear-space}, except that instead of ordering the elements of~$S$ by their probabilities, we order them in increasing order of their working-set numbers $w(i)$. Whenever an element from $D_j$ is reported, we move the element to $D_1$ and for $k = 1,2,\ldots,j-1$ shift one element from $D_k$ to $D_{k+1}$ in order to fill the space left in~$D_j$ while keeping the ordering based on $w(i)$, just as in the working-set structure \cite{unified}. All $2^{2^{j-1}}$ elements in $D_{j-1}$ have been reported more recently than the current element reported from $D_j$, so an analysis similar to \eqref{:equation:bfeq} shows that queries are answered in $\BigOh{\sqrt{\log w(i)}}$ time. The structure uses $\BigOh{n}$ space. 

\begin{theorem}
\label{:theorem:working-set}
Let $w(i)$ denote the number of distinct predecessors reported since the last time the predecessor of $i$ was reported, or $n$ if the predecessor of $i$ has not yet been reported. It is possible to construct a data structure that performs a predecessor search for item $i$ in worst-case time $\BigOh{\sqrt{\log w(i)}}$ using $\BigOh{n}$ space.
\end{theorem}

\section{Conclusion}
\label{:conclusion}

In this paper, we have introduced the idea of biased predecessor search in bounded universes. Two different categories of data structures were considered: one with query times that are logarithmic in the entropy of the query distribution (with space that is a function of $U$), and one with linear space (with query times larger than logarithmic in the entropy). We also considered the cases of individual query times on universe elements with general weights and of query times related to the working-set number.

Our results leave open several possible directions for future research:

\begin{enumerate}
\item Is it possible to achieve $\BigOh{\log H}$ query time and $\BigOh{n}$ space?

\item The reason for desiring a $\BigOh{\log H}$ query time comes from the fact that $H \le \log U$ and the fact that the usual data structures for predecessor searching have query time $\BigOh{\log \log U}$. Of course, this is not optimal: other results have since improved this upper bound \cite{bf,pt}. Is it possible to achieve a query time of, for example, $\BigOh{\log H / \log \log U}$?

\item What lower bounds can be stated in terms of either the input or output entropies? Clearly $\BigOh{U}$ space suffices for $\BigOh{1}$ query time, and so such lower bounds must place restrictions on space usage.
\end{enumerate}

\end{document}